\documentclass[prx,floatfix,unsortedaddress,superscriptaddress]
{revtex4-2}

\usepackage{graphicx}
\usepackage{amsmath} 
\usepackage{amssymb}
\usepackage{siunitx}
\usepackage[colorlinks]{hyperref}
\usepackage{xcolor}

\definecolor{linkcolor}{RGB}{0,83,166}
\hypersetup{
  colorlinks = true,
  allcolors = {linkcolor}
}
\newcommand{\be}{\begin{equation}}
\newcommand{\ee}{\end{equation}}
\newcommand{\ba}{\begin{eqnarray}}
\newcommand{\ea}{\end{eqnarray}}

\begin{document}

\title{Comment on: ``Dynamics of disordered quantum systems with two- and three-dimensional tensor networks'' arXiv:2503.05693}
\date{\today}

\newcommand{\affildw}{D-Wave Quantum Inc., Burnaby, British Columbia, Canada}

\newcommand{\affilsfu}{Department of Physics, Simon Fraser University, Burnaby, British Columbia, Canada}
\newcommand{\affilubc}{Department of Physics and Astronomy and Quantum Matter Institute, University of British Columbia, Vancouver, British Columbia, Canada}
\newcommand{\affiljag}{Jagiellonian University, Institute of Theoretical Physics, {\L}ojasiewicza 11, PL-30348 Krak\'ow, Poland}

\author{Andrew D.~King}
\email[]{aking@dwavesys.com}
\affiliation{\affildw}

\author{Alberto Nocera}
\affiliation{\affilubc}
\author{Marek M. Rams}
\affiliation{\affiljag}
\author{Jacek Dziarmaga}
\affiliation{\affiljag}
\author{Jack Raymond}
\affiliation{\affildw}
\author{Nitin Kaushal}
\affiliation{\affilubc}

\author{Anders W.~Sandvik}
\affiliation{Department of Physics, Boston University, Boston, MA, USA.}

\author{Gonzalo Alvarez}
\affiliation{Computational Sciences and Engineering Division, Oak Ridge National Laboratory, Oak Ridge, TN 37831, USA}

\author{Juan Carrasquilla}
\affiliation{Institute for Theoretical Physics, ETH Z\"urich, 8093, Switzerland}
\affiliation{Vector Institute, MaRS  Centre,  Toronto,  Ontario,  M5G  1M1,  Canada}
\affiliation{Department of Physics and Astronomy, University of Waterloo, Ontario, N2L 3G1, Canada}

\author{Marcel Franz}
\affiliation{\affilubc}

\author{Mohammad H.~Amin}
\email[]{amin@dwavesys.com}
\affiliation{\affildw}
\affiliation{\affilsfu}

\begin{abstract}
In a recent preprint~\cite{Tindall} (arXiv:2503.05693), Tindall~{\it et al.}~presented impressive classical simulations of quantum dynamics using tensor networks.  Their methods represent a significant improvement in the classical state of the art, and in some cases show lower errors than recent simulations of quantum dynamics using a quantum annealer~\cite{King2025}.  However, of the simulations in Ref.~\cite{King2025}, Ref.~\cite{Tindall} did not attempt the most complex lattice geometry, nor reproduce the largest simulations in 3D lattices, nor simulate the longest simulation times, nor simulate the low-precision ensembles in which correlations grow the fastest, nor produce the full-state and fourth-order observables produced in Ref.~\cite{King2025}.  Thus this work should not be misinterpreted as having overturned the claim of Ref.~\cite{King2025}: the demonstration of quantum simulations beyond the reach of classical methods.  Rather, these classical advances narrow the parameter space in which beyond-classical computation has been demonstrated.  In the near future these classical methods can be combined with quantum simulations to help sharpen the boundary between classical and quantum simulability.
\end{abstract}

\maketitle

Tindall {\em et al.}~\cite{Tindall} used a belief propagation~(BP) tensor-network technique~\cite{Alkabetz2021} to simulate the coherent quantum annealing dynamics of locally interacting 2D and 3D spin glasses, with some of the same topologies and parameters used in Ref.~\cite{King2025}. They used a simple-update~\cite{Jiang2008} scheme to evolve the many-body tensor-network wave function via BP~\cite{Tindall2023}, while incorporating two sophisticated variants of BP methods to compute one- and two-point expectation values, based specifically on recently developed matrix product state (MPS) message passing contractions~\cite{Guo2023} and loop series expansion algorithms~\cite{Evenbly2024}.

These advances are a welcome addition to the portfolio of classical approaches that one can use to simulate quantum dynamics, and some of their results go beyond what we were able to achieve classically, notably improving upon the PEPS results shown in Ref.~\cite{King2025}.  These methods are particularly useful for fast quenches, where correlations remain short-ranged and treelike.  As correlation length grows, longer-range loop corrections are needed, at exponentially growing cost.

In Ref.~\cite{King2025}, we used a D-Wave Advantage2$^{\rm TM}$ QPU to simulate a wide range of problems, parameters, and calculated quantities:

\begin{itemize}
    \item {\bf Five different geometries:} 2D square lattices, 3D diamond lattices, 3D cubic-dimer lattices, 3D cubic-nodimer lattices, and dimerized biclique graphs. 
    
    \item {\bf Two sets of parameters:} ``high-precision problems,'' where coupling constants were randomly selected as $\frac{a}{128}$ with $a\in \mathbb{Z}$ in the interval $\{-128,...,128\}$, and ``low-precision problems,'' where couplers were randomly chosen from $\{-1,1\}$ (the widely-studied $\pm J$ model).  Due mostly to the larger average coupling energy, correlations spread faster in the latter.

    \item {\bf Physical quantities:} We perform fair sampling from the QPU, which allows inference of any many-body physical quantity from the quenched distribution of classical spin states. In particular, Edwards-Anderson spin-glass order parameter (two-point correlations) and Binder cumulant (four-point correlations) were used to validate expected behavior at large scale.

    \item {\bf Annealing times:} The QPU solved problems with annealing times $t_a$ ranging from $\SI{7}{ns}$ to $\SI{40}{ns}$, demonstrating agreement with quantum mechanics through the correct critical exponents based on hundreds to thousands of spin-glass instances in each parameterization.  We did not compute MPS ground truths for $t_a>\SI{20}{ns}$, but it is feasible to do so for small inputs with reasonable HPC resources.

    \item {\bf Problem sizes:} Simulations were demonstrated on up to 567 qubits, yielding universal critical exponents consistent with literature values across all geometries.
\end{itemize}

\begin{figure}
    \centering
    \includegraphics[scale=1]{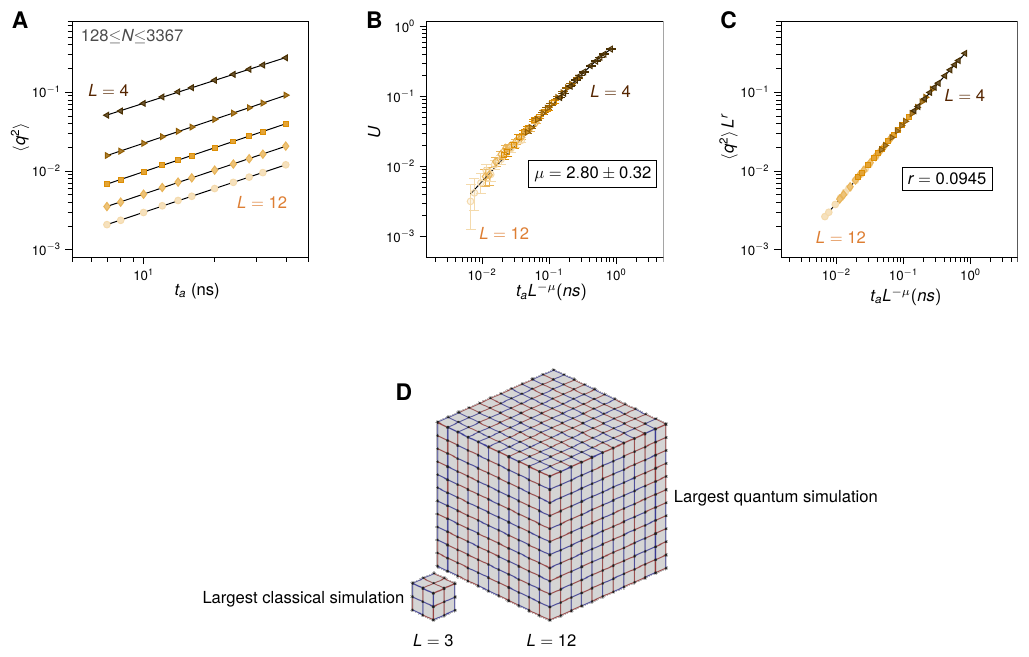}
    \caption{{\bf Dynamic scaling analysis of 3D-dimer $\pm J$ spin glasses on up to $N=3367$ qubits}. {\bf A}: Edwards-Anderson order parameter $\langle q^2\rangle$ computed from $O(N^2)$ two-body correlations for 300 to 3900 input problems per size, with power-law fit lines consistent with quantum critical dynamics up to thousands of qubits and up to $\SI{40}{ns}$ evolution time.  {\bf B}: Binder cumulant $U$ computed from $O(N^4)$ two- and four-body correlations; horizontal collapse yields an estimated Kibble-Zurek exponent $\mu = 2.80\pm 0.32$, very close to the literature value of $2.85$.  {\bf C}: Finite-size dynamic collapse of $\langle q^2\rangle$ based on horizontal rescaling $L^{-\mu}$ and a nontrivial vertical rescaling by $L^r$~\cite{King2023}.  {\bf D}: Depiction of largest lattices simulated with tensor networks in Ref.~\cite{Tindall} ($L{=}3$) and simulated with a quantum annealer here ($L{=}12$).}
    \label{fig:1}
\end{figure}

Tindall {\em et al.}~reproduced only a small subset of our results, as presented in Ref.~\cite{Tindall}. Below, we provide a more detailed, itemized explanation:

\begin{enumerate}
\item Biclique problems are not tractable using the techniques discussed in their paper~\cite{Tindall} or any methods that rely on local interactions, and no results were presented.

\item For other problem types, the paper presented extensive results primarily for 2D lattices. The Kibble-Zurek exponent was estimated based on a small subset of $O(N)$ two-point correlators on two lattice sizes (14$\times$14 and 18$\times$18). While this represents an impressive extension of the state of the art, we deliberately did not claim that 2D lattices were beyond the reach of classical~\cite{King2025}, as low-dimensional systems are easier to simulate classically.

\item Even for 2D problems, only two-point correlations are estimated, and not the Binder cumulant, which requires four-point correlations.  The challenge initially posed in our paper~\cite{King2025} was to sample full states from the post-quench distribution; this was not done, so a comparison of the final distribution against a ground truth, such as provided in our work, was not possible.  In addition to sufficient bond dimension, correctness of the 2D method in the final contraction (measurement) requires correlation length much less than system size; higher order statistics and full sampling are more susceptible to errors relative to two-point correlators.
  
\item For 3D lattices, their demonstrations were limited to systems with up to 54 qubits, which is comparable to what we achieved using TDVP-MPS~\cite{Haegeman2011,Haegeman2016}.  Observed errors in their method increase as a function of system size; despite this, they predict that the errors will decrease for larger sizes due to the disappearance of the loops of length $L_z=3$ in the periodic dimension.  This expectation is inconsistent with our own results (Fig.~S37A~\cite{King2025}) which indicate that boundary conditions remain important up to larger sizes, particularly for quench times $t_a \geq \SI{20}{ns}$.  Given model variation and significant differences in the contraction method in 2D compared to 3D, it would be useful to demonstrate results in 3D beyond the 3$\times$3$\times$3 dimer case.  Moreover, a dynamic finite-size scaling collapse of the Binder cumulant should be shown to verify correct quantum critical scaling at large scales.

\item At their 50-qubit diamond example, computing each two-point correlation takes 15.5 CPU-seconds; naively extending this timing to the full set of four-point correlations on the largest instance we simulated---567 qubits---suggests that it would take hundreds of thousands of CPU-years to estimate the Binder cumulant of a 300-instance ensemble using the standard formula, as we did.  There may be faster estimation methods, for example based on sampling, but the point is that polynomial scaling does not imply practicality.

\item Fig.~4 of their paper suggests that TEBD-MPS method can be used as tool to obtain ground truths (evidenced by a single nearest-neighbor correlation).  In our work we found that the TDVP-MPS algorithm is the best approach to finding, and that the TEBD method, while simpler to implement, it is not a feasible route to provide ground truths at medium to large scales. We also point out that the loop-corrected BP algorithm developed in Ref.~\cite{Evenbly2024} must in principle be compared against converged TDVP-MPS evaluations or PEPS evaluations using a boundary MPS contraction whose bond dimension has been chosen large enough such that any truncation errors are negligible and the results can be considered numerically exact. This explicit benchmark---involving the increase of the $\chi_{\text{BP}}$ and $l_{max}$ parameters---against TDVP-MPS or large-scale PEPS has not been performed.

\item They studied only ``high-precision'' problems, whereas ``low-precision'' problems are significantly more challenging for their loop correction method due to the faster growth of correlations. Given this, we expect their method would struggle to handle low-precision inputs at $\SI{20}{ns}$.
  
\item Our Binder cumulant calculations using QPU extend up to $\SI{40}{ns}$ of annealing time, far beyond their demonstrations.

\item After the appearance of Ref.~\cite{Tindall}, we extended our 3D-dimer simulations from a maximum size of $432$ qubits to $3367$ qubits ($L=12$) on a newer Advantage2 processor, using ensembles of 300 to 3900 instances depending on problem size (Fig.~\ref{fig:1}), allowing an improved estimate of the Kibble-Zurek exponent with diminished finite-size effects.  As classical simulations improve, so too do quantum simulations.

\item The geometries studied in Ref.~\cite{King2025} were chosen specifically because literature values for their universal critical exponents existed, allowing us to verify results at large sizes. However, our QPUs can solve a much broader range of problem geometries, many of which remain inaccessible to the techniques used in Ref.~\cite{Tindall}.  Designing adversarial lattices to make their methods fail would be easy with the addition of unstructured short cycles.

\end{enumerate}

We now also address recent results of Mauron and Carleo~\cite{Carleo}, who used a fourth-order Jastrow ansatz to simulate quantum dynamics of diamond lattices of size $N\leq 128$ at the shortest annealing time of $t_a=\SI{7}{ns}$.

\begin{enumerate}
    \item Like Tindall {\em et al.}, they appear to have advanced the state of the art in classical simulations of these dynamics in their respective approach.  However this was only demonstrated in high-precision diamond lattices at $t_a>\SI{7}{ns}$, where correlations are rather short-ranged. 
    \item They performed a two-dimensional scaling in system size, keeping $L_z$ constant, and extrapolate a relation between simulation error (correlation error) and TDVP error $\overline{R^2}$.  We remark that the linear relation assumed and plotted in Fig.~4a is no more justified by the data than would be a quadratic or even exponential relation.  They also assume, without evidence, that this relation would hold when they change from two-dimensional scaling of system size to three-dimensional scaling.
    \item They further went on to present and compare against extrapolations of computational cost for $t_a=\SI{7}{ns}$.  We did not present the same extrapolation in our paper because we considered the correlations in these simulations to be too short-ranged and therefore vulnerable to local methods such as shown here.
    \item Resource costs for their Monte Carlo procedure were not considered in their analysis.
    \item Their paper did not attempt to handle square, 3D-dimer, or biclique simulations, or any simulations longer than $\SI{7}{ns}$.  In other words, out of all the simulations performed in our work, they only showed results on the sparsest example with the shortest and most treelike correlations.
\end{enumerate}

We emphasize that the above points are not meant to diminish the remarkable achievements of Tindall {\em et al.}~\cite{Tindall}, or Mauron and Carleo~\cite{Carleo}, or suggest that further advancements in numerical simulations are impossible. We welcome focused research on leading classical methods that are critical in helping identify where quantum computing technologies perform calculations beyond the reach of classical technology. We are merely pointing out that the evidence at hand does not refute our conclusions on the beyond-classical capabilities of D-Wave Advantage2 processors.

\bibliography{response}

\begin{thebibliography}{11}%
\makeatletter
\providecommand \@ifxundefined [1]{%
 \@ifx{#1\undefined}
}%
\providecommand \@ifnum [1]{%
 \ifnum #1\expandafter \@firstoftwo
 \else \expandafter \@secondoftwo
 \fi
}%
\providecommand \@ifx [1]{%
 \ifx #1\expandafter \@firstoftwo
 \else \expandafter \@secondoftwo
 \fi
}%
\providecommand \natexlab [1]{#1}%
\providecommand \enquote  [1]{``#1''}%
\providecommand \bibnamefont  [1]{#1}%
\providecommand \bibfnamefont [1]{#1}%
\providecommand \citenamefont [1]{#1}%
\providecommand \href@noop [0]{\@secondoftwo}%
\providecommand \href [0]{\begingroup \@sanitize@url \@href}%
\providecommand \@href[1]{\@@startlink{#1}\@@href}%
\providecommand \@@href[1]{\endgroup#1\@@endlink}%
\providecommand \@sanitize@url [0]{\catcode `\\12\catcode `\$12\catcode
  `\&12\catcode `\#12\catcode `\^12\catcode `\_12\catcode `\%12\relax}%
\providecommand \@@startlink[1]{}%
\providecommand \@@endlink[0]{}%
\providecommand \url  [0]{\begingroup\@sanitize@url \@url }%
\providecommand \@url [1]{\endgroup\@href {#1}{\urlprefix }}%
\providecommand \urlprefix  [0]{URL }%
\providecommand \Eprint [0]{\href }%
\providecommand \doibase [0]{https://doi.org/}%
\providecommand \selectlanguage [0]{\@gobble}%
\providecommand \bibinfo  [0]{\@secondoftwo}%
\providecommand \bibfield  [0]{\@secondoftwo}%
\providecommand \translation [1]{[#1]}%
\providecommand \BibitemOpen [0]{}%
\providecommand \bibitemStop [0]{}%
\providecommand \bibitemNoStop [0]{.\EOS\space}%
\providecommand \EOS [0]{\spacefactor3000\relax}%
\providecommand \BibitemShut  [1]{\csname bibitem#1\endcsname}%
\let\auto@bib@innerbib\@empty
\bibitem [{\citenamefont {Tindall}\ \emph {et~al.}(2025)\citenamefont
  {Tindall}, \citenamefont {Mello}, \citenamefont {Fishman}, \citenamefont
  {Stoudenmire},\ and\ \citenamefont {Sels}}]{Tindall}%
  \BibitemOpen
  \bibfield  {author} {\bibinfo {author} {\bibfnamefont {J.}~\bibnamefont
  {Tindall}}, \bibinfo {author} {\bibfnamefont {A.}~\bibnamefont {Mello}},
  \bibinfo {author} {\bibfnamefont {M.}~\bibnamefont {Fishman}}, \bibinfo
  {author} {\bibfnamefont {M.}~\bibnamefont {Stoudenmire}},\ and\ \bibinfo
  {author} {\bibfnamefont {D.}~\bibnamefont {Sels}},\ }\href
  {https://doi.org/10.48550/ARXIV.2503.05693} {\bibinfo {title} {Dynamics of
  disordered quantum systems with two- and three-dimensional tensor networks}}
  (\bibinfo {year} {2025}),\ \Eprint {https://arxiv.org/abs/2503.05693}
  {arXiv:2503.05693} \BibitemShut {NoStop}%
\bibitem [{\citenamefont {King}\ \emph {et~al.}(2025)\citenamefont {King},
  \citenamefont {Nocera}, \citenamefont {Rams}, \citenamefont {Dziarmaga},
  \citenamefont {Wiersema} \emph {et~al.}}]{King2025}%
  \BibitemOpen
  \bibfield  {author} {\bibinfo {author} {\bibfnamefont {A.~D.}\ \bibnamefont
  {King}}, \bibinfo {author} {\bibfnamefont {A.}~\bibnamefont {Nocera}},
  \bibinfo {author} {\bibfnamefont {M.~M.}\ \bibnamefont {Rams}}, \bibinfo
  {author} {\bibfnamefont {J.}~\bibnamefont {Dziarmaga}}, \bibinfo {author}
  {\bibfnamefont {R.}~\bibnamefont {Wiersema}}, \emph {et~al.},\ }\bibfield
  {title} {\bibinfo {title} {Beyond-classical computation in quantum
  simulation},\ }\href {https://doi.org/10.1126/science.ado6285} {\bibfield
  {journal} {\bibinfo  {journal} {Science}\ ,\ \bibinfo {pages} {eado6285}}
  (\bibinfo {year} {2025})}\BibitemShut {NoStop}%
\bibitem [{\citenamefont {Mauron}\ and\ \citenamefont {Carleo}(2025)}]{Carleo}%
  \BibitemOpen
  \bibfield  {author} {\bibinfo {author} {\bibfnamefont {L.}~\bibnamefont
  {Mauron}}\ and\ \bibinfo {author} {\bibfnamefont {G.}~\bibnamefont
  {Carleo}},\ }\href {https://doi.org/10.48550/ARXIV.2503.08247} {\bibinfo
  {title} {Challenging the {{Quantum Advantage Frontier}} with {{Large-Scale
  Classical Simulations}} of {{Annealing Dynamics}}}} (\bibinfo {year}
  {2025}),\ \Eprint {https://arxiv.org/abs/2503.08247} {arXiv:2503.08247}
  \BibitemShut {NoStop}%
\bibitem [{\citenamefont {Alkabetz}\ and\ \citenamefont
  {Arad}(2021)}]{Alkabetz2021}%
  \BibitemOpen
  \bibfield  {author} {\bibinfo {author} {\bibfnamefont {R.}~\bibnamefont
  {Alkabetz}}\ and\ \bibinfo {author} {\bibfnamefont {I.}~\bibnamefont
  {Arad}},\ }\bibfield  {title} {\bibinfo {title} {Tensor networks contraction
  and the belief propagation algorithm},\ }\href
  {https://doi.org/10.1103/PhysRevResearch.3.023073} {\bibfield  {journal}
  {\bibinfo  {journal} {Physical Review Research}\ }\textbf {\bibinfo {volume}
  {3}},\ \bibinfo {pages} {023073} (\bibinfo {year} {2021})}\BibitemShut
  {NoStop}%
\bibitem [{\citenamefont {Jiang}\ \emph {et~al.}(2008)\citenamefont {Jiang},
  \citenamefont {Weng},\ and\ \citenamefont {Xiang}}]{Jiang2008}%
  \BibitemOpen
  \bibfield  {author} {\bibinfo {author} {\bibfnamefont {H.~C.}\ \bibnamefont
  {Jiang}}, \bibinfo {author} {\bibfnamefont {Z.~Y.}\ \bibnamefont {Weng}},\
  and\ \bibinfo {author} {\bibfnamefont {T.}~\bibnamefont {Xiang}},\ }\bibfield
   {title} {\bibinfo {title} {Accurate {{Determination}} of {{Tensor Network
  State}} of {{Quantum Lattice Models}} in {{Two Dimensions}}},\ }\href
  {https://doi.org/10.1103/PhysRevLett.101.090603} {\bibfield  {journal}
  {\bibinfo  {journal} {Physical Review Letters}\ }\textbf {\bibinfo {volume}
  {101}},\ \bibinfo {pages} {090603} (\bibinfo {year} {2008})}\BibitemShut
  {NoStop}%
\bibitem [{\citenamefont {Tindall}\ and\ \citenamefont
  {Fishman}(2023)}]{Tindall2023}%
  \BibitemOpen
  \bibfield  {author} {\bibinfo {author} {\bibfnamefont {J.}~\bibnamefont
  {Tindall}}\ and\ \bibinfo {author} {\bibfnamefont {M.}~\bibnamefont
  {Fishman}},\ }\bibfield  {title} {\bibinfo {title} {Gauging tensor networks
  with belief propagation},\ }\href
  {https://doi.org/10.21468/SciPostPhys.15.6.222} {\bibfield  {journal}
  {\bibinfo  {journal} {SciPost Physics}\ }\textbf {\bibinfo {volume} {15}},\
  \bibinfo {pages} {222} (\bibinfo {year} {2023})}\BibitemShut {NoStop}%
\bibitem [{\citenamefont {Guo}\ \emph {et~al.}(2023)\citenamefont {Guo},
  \citenamefont {Poletti},\ and\ \citenamefont {Arad}}]{Guo2023}%
  \BibitemOpen
  \bibfield  {author} {\bibinfo {author} {\bibfnamefont {C.}~\bibnamefont
  {Guo}}, \bibinfo {author} {\bibfnamefont {D.}~\bibnamefont {Poletti}},\ and\
  \bibinfo {author} {\bibfnamefont {I.}~\bibnamefont {Arad}},\ }\bibfield
  {title} {\bibinfo {title} {Block belief propagation algorithm for
  two-dimensional tensor networks},\ }\href
  {https://doi.org/10.1103/PhysRevB.108.125111} {\bibfield  {journal} {\bibinfo
   {journal} {Physical Review B}\ }\textbf {\bibinfo {volume} {108}},\ \bibinfo
  {pages} {125111} (\bibinfo {year} {2023})}\BibitemShut {NoStop}%
\bibitem [{\citenamefont {Evenbly}\ \emph {et~al.}(2024)\citenamefont
  {Evenbly}, \citenamefont {Pancotti}, \citenamefont {Milsted}, \citenamefont
  {Gray},\ and\ \citenamefont {Chan}}]{Evenbly2024}%
  \BibitemOpen
  \bibfield  {author} {\bibinfo {author} {\bibfnamefont {G.}~\bibnamefont
  {Evenbly}}, \bibinfo {author} {\bibfnamefont {N.}~\bibnamefont {Pancotti}},
  \bibinfo {author} {\bibfnamefont {A.}~\bibnamefont {Milsted}}, \bibinfo
  {author} {\bibfnamefont {J.}~\bibnamefont {Gray}},\ and\ \bibinfo {author}
  {\bibfnamefont {G.~K.-L.}\ \bibnamefont {Chan}},\ }\href
  {https://doi.org/10.48550/ARXIV.2409.03108} {\bibinfo {title} {Loop {{Series
  Expansions}} for {{Tensor Networks}}}} (\bibinfo {year} {2024}),\ \Eprint
  {https://arxiv.org/abs/2409.03108} {arXiv:2409.03108} \BibitemShut {NoStop}%
\bibitem [{\citenamefont {Haegeman}\ \emph {et~al.}(2011)\citenamefont
  {Haegeman}, \citenamefont {Cirac}, \citenamefont {Osborne}, \citenamefont
  {Pi{\v{z}}orn}, \citenamefont {Verschelde},\ and\ \citenamefont
  {Verstraete}}]{Haegeman2011}%
  \BibitemOpen
  \bibfield  {author} {\bibinfo {author} {\bibfnamefont {J.}~\bibnamefont
  {Haegeman}}, \bibinfo {author} {\bibfnamefont {J.~I.}\ \bibnamefont {Cirac}},
  \bibinfo {author} {\bibfnamefont {T.~J.}\ \bibnamefont {Osborne}}, \bibinfo
  {author} {\bibfnamefont {I.}~\bibnamefont {Pi{\v{z}}orn}}, \bibinfo {author}
  {\bibfnamefont {H.}~\bibnamefont {Verschelde}},\ and\ \bibinfo {author}
  {\bibfnamefont {F.}~\bibnamefont {Verstraete}},\ }\bibfield  {title}
  {\bibinfo {title} {Time-dependent variational principle for quantum
  lattices},\ }\href {https://doi.org/10.1103/PhysRevLett.107.070601}
  {\bibfield  {journal} {\bibinfo  {journal} {Physical Review Letters}\
  }\textbf {\bibinfo {volume} {107}},\ \bibinfo {pages} {1} (\bibinfo {year}
  {2011})}\BibitemShut {NoStop}%
\bibitem [{\citenamefont {Haegeman}\ \emph {et~al.}(2016)\citenamefont
  {Haegeman}, \citenamefont {Lubich}, \citenamefont {Oseledets}, \citenamefont
  {Vandereycken},\ and\ \citenamefont {Verstraete}}]{Haegeman2016}%
  \BibitemOpen
  \bibfield  {author} {\bibinfo {author} {\bibfnamefont {J.}~\bibnamefont
  {Haegeman}}, \bibinfo {author} {\bibfnamefont {C.}~\bibnamefont {Lubich}},
  \bibinfo {author} {\bibfnamefont {I.}~\bibnamefont {Oseledets}}, \bibinfo
  {author} {\bibfnamefont {B.}~\bibnamefont {Vandereycken}},\ and\ \bibinfo
  {author} {\bibfnamefont {F.}~\bibnamefont {Verstraete}},\ }\bibfield  {title}
  {\bibinfo {title} {Unifying time evolution and optimization with matrix
  product states},\ }\href {https://doi.org/10.1103/PhysRevB.94.165116}
  {\bibfield  {journal} {\bibinfo  {journal} {Physical Review B}\ }\textbf
  {\bibinfo {volume} {94}},\ \bibinfo {pages} {165116} (\bibinfo {year}
  {2016})}\BibitemShut {NoStop}%
\bibitem [{\citenamefont {King}\ \emph {et~al.}(2023)\citenamefont {King},
  \citenamefont {Raymond}, \citenamefont {Lanting}, \citenamefont {Harris},
  \citenamefont {Zucca} \emph {et~al.}}]{King2023}%
  \BibitemOpen
  \bibfield  {author} {\bibinfo {author} {\bibfnamefont {A.~D.}\ \bibnamefont
  {King}}, \bibinfo {author} {\bibfnamefont {J.}~\bibnamefont {Raymond}},
  \bibinfo {author} {\bibfnamefont {T.}~\bibnamefont {Lanting}}, \bibinfo
  {author} {\bibfnamefont {R.}~\bibnamefont {Harris}}, \bibinfo {author}
  {\bibfnamefont {A.}~\bibnamefont {Zucca}}, \emph {et~al.},\ }\bibfield
  {title} {\bibinfo {title} {Quantum critical dynamics in a 5,000-qubit
  programmable spin glass},\ }\href
  {https://doi.org/10.1038/s41586-023-05867-2} {\bibfield  {journal} {\bibinfo
  {journal} {Nature}\ }\textbf {\bibinfo {volume} {617}},\ \bibinfo {pages}
  {61} (\bibinfo {year} {2023})}\BibitemShut {NoStop}%
\end{thebibliography}%

 \end{document}